\documentclass[preprint,superscriptaddress,aip,jcp]{revtex4-1}
\usepackage{graphicx}
\usepackage{epstopdf}
\usepackage{float}
\usepackage{bm}
\usepackage[utf8]{inputenc}
\usepackage{color}
\begin{document}
\title{Three-dimensional patchy lattice model: ring formation and phase separation}

\author{J. M. Tavares}
\affiliation{ Centro de F{\'\i}sica Te\'orica e Computacional, Universidade de Lisboa,
Avenida Professor Gama Pinto 2, P-1649-003 Lisbon, Portugal and
Instituto Superior de Engenharia de Lisboa, Rua Conselheiro Em\'{\i}dio Navarro 1, 
P-1950-062 Lisbon, Portugal}

\author{N. G. Almarza}
\affiliation{Instituto de Qu{\'\i}mica F{\'\i}sica Rocasolano, CSIC, Serrano 119, E-28006 Madrid, Spain }

\author{M. M. Telo da Gama}
\affiliation{Centro de F{\'\i}sica Te\'orica e Computacional, Universidade de Lisboa,
Avenida Professor Gama Pinto 2, P-1649-003 Lisbon, Portugal and
Departamento de F{\'\i}sica, Faculdade de Ci{\^e}ncias, Universidade de Lisboa,
Campo Grande, P-1749-016 Lisbon, Portugal}

\date{\today}
\begin{abstract}
We investigate the structural and thermodynamic properties of 
a model of particles with $2$ patches of type $A$ and $10$ patches
of type $B$. Particles are placed on the sites of a face centered cubic lattice with 
the patches oriented along the nearest neighbor directions.
The competition between the self-assembly of chains, rings and networks
on the phase diagram is investigated by carrying out a systematic investigation of this class of models, using an 
extension of Wertheim's theory for associating fluids and Monte Carlo numerical simulations.
We varied the ratio $r\equiv\epsilon_{AB}/\epsilon_{AA}$ of the interaction between patches $A$ and $B$, $\epsilon_{AB}$,
and between $A$ patches, $\epsilon_{AA}$ ($\epsilon_{BB}$ is set to $0$) as well as the relative position of the $A$ patches, 
i.e., the angle $\theta$ between the (lattice) directions of the $A$ patches. 
We found that both $r$ and $\theta$ ($60^\circ,90^\circ,$ or $120^\circ$) have a profound effect on the phase diagram. In 
the empty fluid regime ($r < 1/2$) the phase diagram is re-entrant with a closed miscibility loop. The region around the lower critical point exhibits unusual structural and thermodynamic behavior determined by the presence of relatively short rings. 
The agreement between the results of theory and simulation is excellent for $\theta=120^\circ$ but deteriorates as $\theta$ decreases, revealing the need for new theoretical approaches to describe the structure and thermodynamics of systems dominated by small rings.

\end{abstract}
\maketitle

\section{Introduction}

The microscopic mechanism driving the liquid vapor condensation, characteristic of simple fluids, is the balance between 
isotropic short range repulsions and long range attractions that promote the formation of isotropic clusters of particles, which ultimately become macroscopic at the (critical) phase transition. The addition of short range directional attractions promotes the association of the particles in non spherical clusters that have a profound effect on the phase transition. 
These bonding interactions are designed in recently fabricated nanometer-to-micrometer sized particles, which self-assemble 
into a variety of complex structures \cite{glotzer2005,pawar2010,bianchi2011,pine2012}. By patterning the surface of these particles, control over the anisotropic interactions may be achieved and the structure of the self-assembled clusters 
becomes tunable. The formation of (large) anisotropic clusters competes with (and eventually inhibits) the isotropic phase transition. 
Thus, the design of patchy particles that self-assemble in predefined structures, requires the knowledge of their macroscopic properties.

Indeed, recent studies of the structural and thermodynamic properties of models of patchy particles have shown that the interplay between self-assembly and phase separation leads to unusual macroscopic phases such as empty liquids, optimal networks, equilibrium gels, micellar phases, etc. \cite{bianchi2006,sciortino2009,zaccarelli2007}. 

Patchy particle models with dissimilar patches (of types $A$ and $B$) were introduced in this context \cite{tavares2009a,tavares2009b}, and studied thoroughly, using both theory and simulation, in the case where linear self-assembly dominates \cite{russo2011a,russo2011b,almarza2011,almarza2012}. 
The particles are endowed with two (strong) patches $A$ and one or several (weak) patches $B$. Whenever two patches (one of type $\alpha$ and another of type $\beta$) belonging to neighboring particles overlap (or align), a bond $\alpha\beta$ is formed and the internal energy decreases by $\epsilon_{\alpha\beta}$. By setting $\epsilon_{BB}=0$, only $AA$ and $AB$ bonds may form and for systems with $r\equiv\epsilon_{AB}/\epsilon_{AA}<1/2$, the assembly of linear assemblies of particles connected by consecutive $AA$ bonds (chains or rings) becomes energetically favored \cite{tavares2009b,russo2011a}.
If the $A$ patches are placed on opposite sides of the particle, ring formation is negligible (or impossible in lattice models) and the phase behavior is dictated by a peculiar balance between the entropic gain and the energetic cost of forming $AB$ bonds (compared to $AA$ bonds) \cite{russo2011b,almarza2012}. For a given entropic gain, there is a threshold in the energy cost above which no phase transition occurs. Below this threshold, a phase transition between a gas of $AA$ chains (rich in non bonded $A$ patches, or ends) and a liquid formed by a network of long $AA$ chains connected by $AB$ bonds or junctions (and thus rich in junctions) occurs. The liquid binodal of this phase separation is reentrant: on cooling, the density of the liquid at coexistence decreases and approaches that of the vapor.
Remarkably, the results predicted theoretically by Wertheim's first order perturbation theory have been confirmed quantitatively by simulations of both off-lattice 
and lattice models \cite{russo2011b,almarza2011,almarza2012}.    

More recently, the phase behavior of variants of this model, which include more or less pronounced ring formation were investigated \cite{almarza2012b,rovigatti2013}. Using Monte Carlo simulations, it was found that the phase behavior on 2D triangular lattices, changes dramatically with the relative position of the $A$ patches: (i) when the $A$ patches are on opposite lattice directions, preventing ring formation, a reentrant phase diagram, as described above, is obtained in line with \cite{russo2011a,almarza2011}; (ii) when the $A$ patches are on consecutive lattice directions (at $60^\circ$) short rings are formed and no phase transition is observed; (iii) finally, when the $A$ patches are on lattice directions at $120^\circ$, a phase diagram with both an upper and a lower critical point is found. The existence of a closed miscibility loop was confirmed by \cite{rovigatti2013}, for an off-lattice model both theoretically and by computer simulations. Close to the upper critical point the coexistence is between a vapor rich in ends and a liquid rich in junctions; on cooling both phases become reentrant (i.e. the vapor density increases and the liquid density decreases) and become identical at a lower critical point. Near this critical point, the vapor consists (mostly) of isolated rings while the liquid is a network of rings and chains. Moreover, the vapor has a lower energy and a lower entropy than the liquid.

In this paper we carry out a systematic investigation of the phase behavior of patchy particle fluids and their equilibrium structure, i.e., the distribution of self-assembled chains, rings and networks. To that effect, we consider particles with two strong $A$ patches and ten weak $B$ patches, which is a generalization of the model studied in 
\cite{almarza2012}: particles are placed on the sites of a face centered cubic (fcc) lattice with the patches aligned in the nearest neighbor (NN) directions. In the general model the position of the $A$ patches may be varied from $180^\circ$ to $60^\circ$. The phase diagram is calculated, using theory and simulation, for several values of the model parameters. The comparison between the results provides a stringent test of the accuracy of Wertheim's theory in accounting for ring formation.   

The remainder of this paper is arranged as follows: in Sec. \ref{sec:3DModel} we describe the model; in Sec. \ref{sec:Simulation} we describe the Monte Carlo techniques used to compute the phase diagrams; in Sec. \ref{sec:theory}
we carry out the theoretical analysis; in Sec. \ref{sec:results} we compare the results obtained by theory and simulations; finally, in \ref{sec:Conclusions} we discuss these and previous results and perspectives for future work.

\section{Three dimensional model}
\label{sec:3DModel}

\begin{figure} 
\includegraphics[width=150mm,clip=]{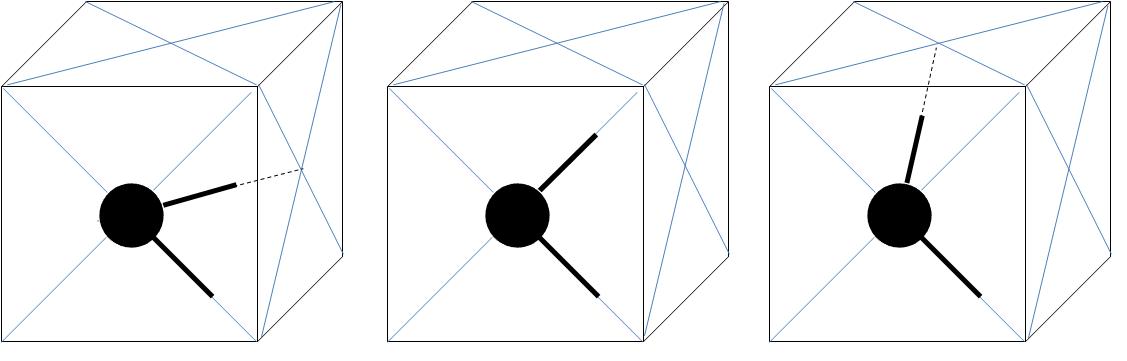}
\caption{Schematic representation of the model. Each cube is a unit cell of an fcc lattice. The disk represents one particle at a lattice site. The thick segments represent the two $A$ patches pointing to nearest neighbors along the lattice directions. From left to right, the angle $\theta$ between the two $A$ patches is $60^\circ$, $90^\circ$, and $120^\circ$. 
The $B$ patches are not represented for clarity: they are segments oriented along the ten remaining nearest neighbor directions.}
\label{fig:model}
\end{figure}

We consider an fcc lattice. Sites on the lattice can be either empty or occupied by, at most, one particle. The particles carry two patches of type $A$, and $10$ of type $B$. Each patch on a particle is oriented in one of the twelve different directions linking the site with its nearest neighbors. The angle $\theta$ between the two $A$ patches can take the values $60^\circ, 90^\circ$ or $120^\circ$ - see figure \ref{fig:model}. The model with $\theta=180^\circ$ was considered in \cite{almarza2012}.

The interaction energy of two particles, $1$ and $2$, is $-\epsilon_{\alpha\beta}<0$ when they are NN on the lattice, and particle 1 has a patch of type $\alpha$ directed towards particle 2, and particle $2$ has a patch of type $\beta$ directed towards particle 1. This configuration will be referred to as an $\alpha\beta$ bond. In all other cases, the interaction energy is 0. 

We set $\epsilon_{BB}=0$, in line with earlier work \cite{russo2011a,russo2011b,almarza2011,almarza2012,almarza2012b,rovigatti2013}, and as a consequence, only $AA$ and $AB$ bonds are formed: $AA$ bonds correspond to linear self-assembly (in chains or rings) and $AB$ bonds to branching points or junctions. 
We choose $r\equiv\epsilon_{AB}/\epsilon_{AA}<1/2$ to favor energetically the formation of chains and rings; the number of 
$B$ patches corresponds to the available volume to form junctions, which are, as a consequence, entropically favorable \cite{russo2011b,almarza2012}. 
When $\theta=180^\circ$ rings are absent, and the competition for self-assembly of chains and junctions drives the re-entrant phase behavior described in the introduction and obtained in \cite{russo2011b,almarza2012}.
Here we consider $\theta \ne 180^\circ$, and self-assembly of rings (defined as sequences of particles connected by $AA$ bonds) is expected to occur with impact on the phase behavior. Chains are energetically unfavored and entropically favored relative to rings of the same size. Ring formation will also affect the formation of junctions: when chains close to form rings, the $A$ patches saturate, and become unavailable to form $AB$ bonds. 

The model under study is controlled by two parameters: $r\equiv\epsilon_{AB}/\epsilon_{AA}$, which sets 
the energetic cost of forming junctions or $AB$ bonds; and $\theta$, the angle between the directions 
of the $A$ patches, through which ring formation (the entropic cost of rings, to be accurate) is controlled.
In the following, we will compute and compare the phase diagrams obtained for different sets of parameters, $(r,\theta)$: by decreasing $r$ the energy cost of forming junctions is increased; by decreasing $\theta$ the entropic cost of rings is decreased and more and shorter rings are expected to form.
  
\section{Simulation Methods}
\label{sec:Simulation}

We have  carried out Markov Chain Monte Carlo simulations (MC) \cite{frenkel-smit,landau-binder}
on the Grand Canonical Ensemble (GCE) using periodic boundary conditions. The number of lattice sites can be 
written as $M=4L^3$, with $L$ the number of times that the unit cell is replicated in each direction.
A given site, $i$, has $n_s+1$ states: $s_i=0,1,\cdots,n_s$, where $n_s$ is the number of orientations that a particle 
can adopt, point its patches to its NN sites on the lattice; $n_s$ depends on $\theta$. For $\theta=60^{\circ}$ and $\theta=120^{\circ}$ we $n_s=24$, whereas for $\theta=90^{\circ}$ $n_s=12$. The state $s_i=0$ corresponds to the site
being empty.

The GCE simulations include two types of MC steps: single site moves, and multiple site moves using a cluster algorithm.

\subsection{Single-site moves}

A single-site move  consists in updating the state of one site of the system. This is done by picking one site of the system at random (with equal probabilities) and selecting one of its possible states, with probabilities $\alpha(s)$, that take into
account the interactions of the site with its NN on the lattice and the values of the temperature, $T$ and the chemical potential $\mu$:
\begin{equation}
\alpha(s) =  \frac{ p(s)} {\sum_{s'=0}^{n_s} p (s') }
\end{equation}
with $p(s)/p(0)  = \exp \left[ - \beta u_i(s) + \beta \mu \right]$, for $s\ne 0$; $u_i(s)$ is the interaction energy of the particle on site $i$, in state $s$, with the particles on NN sites; $\beta \equiv 1/(k_B T)$ ($k_B$ is Boltzmann's constant). 

\subsection{Cluster algorithm}
The second type of moves works by inserting or deleting sequences of particles linked by $AA$ bonds. The algorithm is based on the cluster procedures proposed in our previows work \cite{almarza2012} for the model with $\theta=180^{\circ}$.
In these moves, insertion or deletion is chosen with the same probability, and then the following steps are followed:

\subsubsection{Deletion of particles}

(1) An occupied site is chosen at random; the particle at that site is the {\it root} used to define the cluster of particles 
to be eventually deleted.

(2) One of the $A$ patches of the root is selected to start growing the first branch of the sequence. If this patch is bonded to another $A$ patch of a NN particle, the latter is linked to the root (added to the cluster) with probability $b$ (not linked with probability $1-b$). If there are no $AA$ bonds the particles are not linked. For links established in the previous step, we consider the particles added to the cluster, and check if they participate in any other $AA$ bonds, applying the same probabilistic criterion to incorporate new particles to the cluster. The process stops when no new links are established or when the cluster closes on itself forming a ring.

(3) The growth process (2) is repeated for the second branch (through the second $A$ patch of the root particle).

\subsubsection{Insertion of particles}

(1) A {\it root} particle is inserted in an empty position chosen at random, its orientation chosen from
the $n_s$ possible states at random with equal probabilities.

(2) One of the $A$ patches is chosen at random and then one start growing the cluster at this point,
If the corresponding NN lattice position is empty then one continues the cluster growing
with probability $b$ (or stop it with probability $1-b$)  by inserting a particle with one of its $q_s$ orientations that produce
an $AA$ bond with the previous particle ($q_s=2$ for $\theta=90^{\circ}$; and $q_s=4$ for
both $\theta=60^{\circ}$ and $\theta=120^{\circ}$). The growing process through this branch
finishes either by the probabilistic rule given above or because the NN site to which
the $A$ patch points to is occupied.

(3) Once the growth of the first branch finishes, the same procedure is applied from the
second $A$ patch of the root particle.

\subsubsection{Acceptance criteria}

Taking into account super-detailed balance\cite{frenkel-smit} it is relatively simple
to write down the ratio between the acceptance probabilities
of MC moves involving two configurations that can interconvert, by inserting / removing
a chain of $\Delta N$ particles connected through $AA$ bonds. Let $N_0$ and $N_1$ be
the number of particles in those configurations, with $N_1 = N_0 + \Delta N$.
By taking into account the probabilities of generating
a given trial cluster in both the insertion and deletion attempts we obtain:
\begin{equation}
\frac{ A ( N_1 | N_0) }
     { A ( N_0 | N_1 ) } = e^{ - \beta \left( \Delta U - \mu \Delta N \right)} q_s^{\Delta N -1}
\frac{ (M-N_0) n_s }{N_1} \frac
{ \left( 1 - b \right)^{n_r(N_1)} }
{ \left( 1 - b \right)^{n_r(N_0)} };
\end{equation}
where $A(N'|N)$ is the MC acceptance probability of  the  configuration with $N'$ particles when
created as a trial configuration from the configuration with $N$ particles; $\Delta U$ is
the difference of potential energy between the two configurations, $\Delta U = U(N_1)-U(N_0)$, and 
$n_r(N)$ is the number of possible links rejected by the application of the probabilistic
criterion, during the growth process.

The efficiency of the sampling procedure
will depend on the choice of $b$. We have chosen $b=1-\exp(-\beta \epsilon_{AA}/2)$ by considering the analogies
of the current method with one of the cluster algorithms developed in Ref. \onlinecite{almarza2012}.

The cluster algorithm and its implementation in the numerical codes was checked against
calculations with single-site moves only.
We found that the combination of
the cluster and single-site algorithms improves the efficiency of the method.
Whereas at relatively high temperatures the single site algorithm can sample the phase space accurately, at low 
temperature its performance is rather poor. By contrast, the cluster moves allow an efficient sampling even close to critical points at low temperatures.

\subsubsection{The limit of low temperature}
As in previous work on related models\cite{almarza2011,almarza2012} it is possible to build asymptotic models that capture the limiting behavior at low temperatures, when $r \rightarrow 0.50^-$. 
The adaptation of the cluster algorithm to these models is achieved by forbidding configurations with non-bonded $A$ patches:
the bonding probability is taken to be $b=1$, and deletion moves are rejected if they lead to a non-bonded $A$ patch. The relevant reduced temperature
is then $t^*= k_B T/[(1/2-r)\epsilon_{AA}]$.

\subsection{Computation of the liquid-vapor equilibria}

In order to obtain an overview of the phase diagrams of the systems we used the following procedure: At fixed temperatures, 
we run simulations at different chemical potentials for small systems, $L=6$. From these we obtain an estimate of the region 
in $(\mu,T)$ where liquid-vapor equilibrium is to be found, and an idea of the location of the critical points.

In order to compute the precise location of the liquid-vapor transition
we coupled our GCE simulations with  Thermodynamic Integration
(TI) and Gibbs-Duhem Integration (GDI)\cite{frenkel-smit,almarza2012,almarza2011b}. 
TI is used to obtain, at least, one reference point $(\mu_0,T_0)$ on the liquid-vapor coexistence line, and GDI is used to compute the liquid-vapor binodals using as starting point that computed using TI. Technical details of the procedures can be found elsewhere \cite{almarza2011}. In these calculations we used relatively large system sizes: $L=32$, to
prevent the failure of GDI due to the possibility of {\it jumps} between the liquid and vapor
phases during the integration process. Of course these {\it jumps} are expected to occur in the neighborhood of the critical points, for any system size.

In order to estimate the critical points we use finite-size-scaling techniques \cite{wilding1995}.
The definition of the system-size dependent pseudo-critical properties,
$\mu_c(L)$, $T_c(L)$, $\rho_c(L)$ (critical density) and the procedure to
estimate the critical properties in the thermodynamic limit can be found
in Ref. \onlinecite{almarza2012}.

The pseudo-critical temperatures are obtained as follows: after a rough location of the critical point in the $(\mu,T)$ plane
we run preliminary GCE simulations for a small system size, typically $L=6$, and
apply histogram reweighting techniques\cite{landau-binder,lomba2005} to improve the initial estimates of $T_c(L)$
and $\mu_c(L)$. Once an improved estimate of the critical point is obtained,
we run long simulations, and the results are used after applying histogram
reweighting to compute the pseudo-critical properties for that system size.

The final results for a given system size, $L_0$, are then used to run the preliminary simulations for larger system sizes.
Typically we considered systems up to $L=16$ for $\theta=60^0$, and $\theta=90^0$,
whereas for $\theta=120^0$ we considered systems up to $L=20$ for the upper critical points,
and up to $L=24$ for lower critical points.

\section{Extended Wertheim's theory for $2AnB$ lattice models with rings}
\label{sec:theory}

One of the most successful theories to describe the effects of self-assembly in the thermodynamics of fluids is Wertheim's perturbation theory (WPT) \cite{chapman1988,wertheim1986a}.
In this approximation, the fluid is described as a mixture of species, each corresponding to particles that have a particular set of patches (or bonding sites) bonded. The free energy is obtained using a perturbation theory where the reference system is, usually, the hard sphere fluid, and the interactions that promote bonding are the perturbation, under the following conditions: (i) each patch can take part in one bond only; (ii) two particles can connect to each other through one bond only; (iii) the bonds are independent (and thus loops and rings are neglected). 
The thermodynamic behavior and the equilibrium structure of models of patchy particles have been described successfully using this theory \cite{bianchi2006,russo2011a,almarza2012}. 
For lattice models, like that under study, conditions (i) and (ii) are respected by construction. On the other hand, the independence of the bonds (iii), breaks down, as the lattice introduces spatial correlations between the particles and thus correlations between the bonds. Nevertheless, 
previous works have shown that these correlations may be neglected, as semi-quantitative agreement between the results of WPT and of simulations was observed, under very general conditions \cite{almarza2011,almarza2012}. 
However, WPT fails to account for the phase behaviour obtained by simulations when 
the relative position of the $A$ patches promotes self-assembly of rings \cite{almarza2012b}. To account for the formation of rings, WPT has to be extended \cite{rovigatti2013,sear1994,galindo2002,avlund2011,rovigatti2012}. 
This extension was shown to reproduce accurately the simulation results for an off-lattice model where the particles self-assemble into chains and rings only \cite{rovigatti2012}. More recently, the extension was generalized to systems where junctions are formed \cite{rovigatti2013}. The comparison with simulation revealed that the theory reproduces (at least qualitatively) the simulation results: the existence of two critical points, the re-entrant binodals, the change in the sign of the entropy and energy differences between the coexisting phases, etc. The comparison of the theoretical results with the lattice model simulations, over a wider range of model parameters, provides a much more stringent test of the theory, which as we will show becomes increasingly innacurate as $\theta$ decreases, i.e., as the number of rings increases and their size decreases.  
  
\begin{figure} 
\includegraphics[width=150mm,clip=]{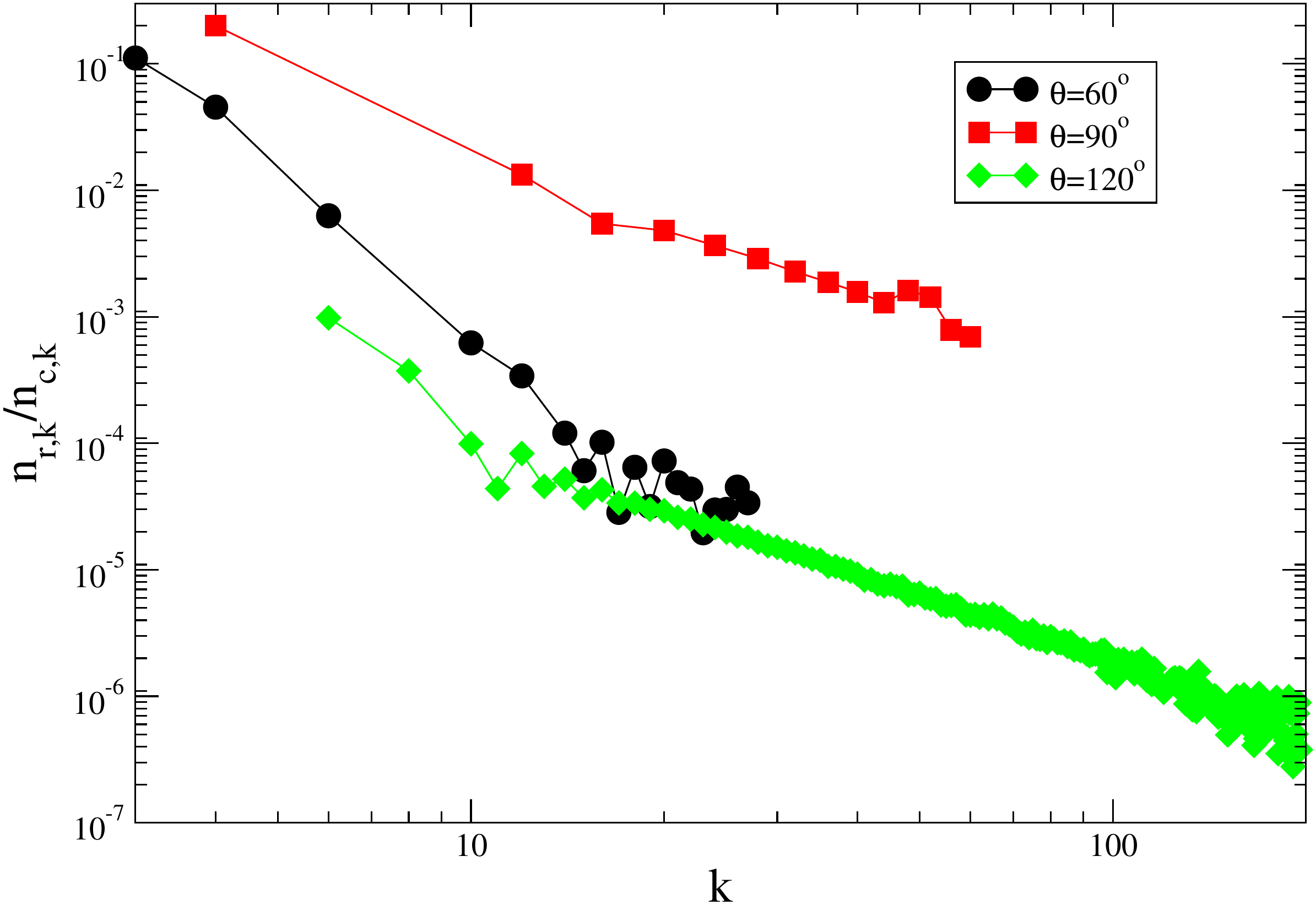}
\caption{Ratio of the number of rings, $n_{r,k}$, and the number of chains, $n_{c,k}$, obtained from a large number ($10^7$ to $10^9$) of realizations of one cluster with $k$ particles (linked by $AA$ bonds), for three values of $\theta$, the angle between the $A$ patches.}
\label{fig:ringdist}
\end{figure}

The free energy per particle of a homogeneous system of particles with 2$A$ and $nB$ patches is,
\begin{equation}
\label{fperturb}
\beta f = \beta f_{ref}+\beta f_b
\end{equation}
where,
\begin{equation}
\label{fref}
\beta f_{ref}=\ln \rho +\frac{1-\rho}{\rho}\ln(1-\rho)
\end{equation}
is the free energy of the reference system (the ideal lattice gas), $\rho$ is the total number density,  
and $f_b$ is the bonding contribution, calculated through the extension of WPT to include rings \cite{rovigatti2013},
\begin{equation}
\label{fb}
f_b=\ln(YX_B^n) -X_A-\frac{n}{2}X_B+\frac{n}{2}+1-\frac{G_0}{\rho}.
\end{equation}
Here, $X_A$ is the fraction of unbonded patches of type $A$, $Y$ is the fraction of particles 
with the two $A$ patches unbonded, and $G_0$ is the number density of rings.
These quantities are related to the number density $\rho$ and to the temperature $T$ 
through the laws of mass action,
\begin{equation}
\label{lma1}
1-\frac{X_A^2}{Y}=\frac{G_1}{\rho},
\end{equation}
\begin{equation}
\label{lma2}
\frac{X_A}{Y}-2\rho\Delta_{AA}X_A-n\rho\Delta_{AB}X_B=1,
\end{equation}
\begin{equation}
\label{lma3}
X_B+2\rho\Delta_{AB}X_AX_B=1.                        
\end{equation}
$\Delta_{AB}$ are integrals of the Mayer functions of two patches $A$ and $B$ on two different particles, 
over their positions and orientations,  
weighted by the pair distribution function of the reference system; for lattice systems \cite{almarza2012},
\begin{equation}
\label{DABlatt}
\Delta_{AB}=v_{b}\left[\exp(\beta\epsilon_{AB})-1\right],
\end{equation}
where $\beta \equiv 1/(k_B T)$ ($k_B$ is Boltzmann's constant), and $v_{b}$ is the volume of a bond. 
In lattice models, this volume is the inverse of the coordination number of the lattice. Therefore, for the present model, $v_b=\frac{1}{12}$. 
$G_i$ is the $i^{th}$ moment of the density distribution of rings,
\begin{equation}
\label{Gi}
G_i=\sum_k k^i W_k (2\rho\Delta_{AA}Y)^k.
\end{equation}
$W_k (2\rho\Delta_{AA}Y)^k$ is the density of rings of size $k$: the formation of one ring requires 
$k$ particles, each with one of two possible orientations; these particles have two unbonded $A$ patches, and their density 
is $\rho Y$; $\Delta_{AA}$ is the probability of forming one $AA$ bond, once the two unbonded $A$ patches are chosen. $W_k$ is the number of configurations of a ring of size $k$. $W_k$ is calculated numerically as in \cite{rovigatti2012,rovigatti2013}: 
linear clusters (i.e. consecutive particles connected by $AA$ bonds) of a given size $k$ are generated in an independent simulation; the number of rings $n_{r,k}$ (i.e. the number of realizations of the cluster with all $A$ patches bonded) and of chains $n_{c,k}$ (i.e. the number of realizations of the cluster with 2 unbonded $A$ patches) is used to calculate  $W_k=n_{r,k}/(2\Delta_{AA} n_{c,k})$. The results depend on the parameter $\theta$ of the model, through $W_k$. The ratio $n_{r,k}/n_{c,k}$ was calculated for $\theta=60^\circ,90^\circ$ and $120^\circ$ (see figure \ref{fig:ringdist}).

Given $W_k(\theta)$, the thermodynamic properties are obtained easily, as the laws of mass action (\ref{lma1},\ref{lma2},\ref{lma3}) yield $X_A$, $X_B$ and $Y$ as a function of $\rho$ and $T$, and (\ref{fperturb},\ref{fref},\ref{fb}) give the free energy as a function of $\rho$, $X_A$, $X_B$ and $Y$.

\section{Results}
\label{sec:results}

Both the theoretical and simulation results indicate the existence of a closed-loop
liquid-vapor equilibrium for the three values of $\theta$ considered in
this work. This happens for a certain range of values: $r^* < r < 1/2$, where
the threshold value $r^*$ depends on $\theta$.
A second relevant conclusion that can be extracted from the finite-size scaling analysis 
of the simulation results concerns the nature of the critical points.
The system-size dependence of the pseudo-critical properties computed
in the simulations was shown to be fully compatible with the three dimensional Ising universality class.
In Tables \ref{table:1} and \ref{table:2} we present some of the results
for the critical points.

\begin{table}
\begin{tabular}{llll|llllll|lll}
\hline
$\theta$  & & $r$ & & $T_c^*$ & & $\rho_c$ & & $\mu_c^*$ & & $t'_c$ & &  $\mu'_c$  \\
\hline
120$^{\circ}$ & & 0.32 & & 0.09174(10) & & 0.0914(5) & & -1.1447(2) & & 0.5096(6) & & -0.804(1)  \\
120$^{\circ}$ & &    0.325 & & 0.08317(6) & & 0.0795(6) & & -1.1275(2) & & 0.4753(4) & & -0.728(1) \\
120$^{\circ}$ & &  0.33  & & 0.07751(4) & & 0.0717(3) & & -1.1170(1) & & 0.4559(3) & & -0.6880(5) \\
120$^{\circ}$ & & 0.35  & & 0.06343(3) & & 0.0597(5) & & -1.09328(6) & & 0.4229(5) & & -0.6219(4) \\
120$^{\circ}$ & & 0.40  & & 0.04093(3) & & 0.0552(3) & & -1.05960(5) & & 0.4093(3) & & -0.5960(5) \\
120$^{\circ}$ & & 0.45  & & 0.02047(2) & & 0.0552(4) & & -1.02980(3) & & 0.4094(4) & & -0.5960(6) \\
120$^{\circ}$ & & 0.5$0^-$  & &            & & 0.0553(3) & &    & & 0.4094(3) & & -0.5961(6) \\
\hline
 90$^{\circ}$ & & 0.405 & & 0.1255 (4) & & 0.278(1) & & -1.2083(9) & & 1.321(5) & & -2.193(10)  \\
 90$^{\circ}$ & & 0.41  & & 0.1106 (2) & & 0.280(1) & & -1.1759(4) & & 1.228(2) & & -1.955(5)  \\
 90$^{\circ}$ & & 0.425 & & 0.08584(8) & & 0.282(1) & & -1.1307(2) & & 1.145(1) & & -1.742(3)  \\
 90$^{\circ}$ & & 0.45  & & 0.05581(5) & & 0.284(1) & & -1.0835(2) & & 1.116(1) & & -1.670(3)  \\
 90$^{\circ}$ & & 0.475 & & 0.02789(3) & & 0.283(1) & & -1.0417(1) & & 1.116(1) & & -1.669(3)  \\
 90$^{\circ}$ & & 0.5$0^-$  & &        & & 0.283(1) & &            & & 1.115(1) & & -1.669(2) \\
\hline
 60$^{\circ}$ & & 0.478 & & 0.0949(12) & & 0.296(2) & & -1.286(4) & & 4.31(6) & & -12.8(2)  \\
 60$^{\circ}$ & & 0.48  & & 0.0798(5) & & 0.296(2) & & -1.238(2) & & 3.99(3) & & -11.9(2)  \\
 60$^{\circ}$ & & 0.485 & & 0.0576(3) & & 0.296(2) & & -1.171(2) & & 3.84(2) & & -11.4(2)  \\
 60$^{\circ}$ & & 0.490 & & 0.0382(2) & & 0.296(2) & & -1.1131(7) & & 3.82(2) & & -11.3(1)  \\
 60$^{\circ}$ & & 0.5$0^-$  & &        & & 0.297(2) & &            & & 3.82(2) & & -11.3(1) \\
\hline
\end{tabular}
\caption{Estimates of the lower critical points for different values of $\theta$ and $r$.
$T_c^*=k_BT_c/\epsilon_{AA}$, $\mu_c^*=\mu_c/\epsilon_{AA}$.
The reduced values $t'_c$ and $\mu'_c$ used to check the asymptotic behavior
are defined as: $t'_c = T^*_c/(1/2-r)$ and $\mu'_c = (1 + \mu^*_c) /(1/2-r)$.  }
\label{table:1}
\end{table}

\begin{table}
\begin{tabular}{lllllll|llllllll}
\hline
$\theta$  & & $r$ & & $T_c^*$ & & $\rho_c$ & & $\theta$ & & r & & $T_c^*$ & & $\rho_c^*$ \\
\hline
120$^{\circ}$ & & 0.32 & & 0.1231(2) & & 0.150(1) & & 120$^{\circ}$ & & 0.40 & & 0.2163(1) & & 0.280(1) \\
120$^{\circ}$ & & 0.50 & & 0.2913(1) & & 0.323(1) & &  90$^{\circ}$ & & 0.405  & & 0.1514(3) & & 0.277(1) \\
 90$^{\circ}$ & & 0.45 & & 0.2184(1) & & 0.294(1) & &  90$^{\circ}$ & & 0.50   & & 0.2607(1) & & 0.308(1) \\
 60$^{\circ}$ & & 0.478& & 0.108 (2) & & 0.295(2) & &  60$^{\circ}$ & & 0.48   & & 0.1210(6) & & 0.293(2) \\
 60$^{\circ}$ & & 0.49 & & 0.1468(4) & & 0.291(2) & &  60$^{\circ}$ & & 0.48   & & 0.1618(3) & & 0.290(2) \\
\hline
\end{tabular}
\caption{Estimates of the upper critical points for selected values of $\theta$ and $r$.
$T_c^*=k_BT_c/\epsilon_{AA}$.}
\label{table:2}
\end{table}

\subsection{$\theta=120^\circ$}

\begin{figure} 
\includegraphics[width=150mm,clip=]{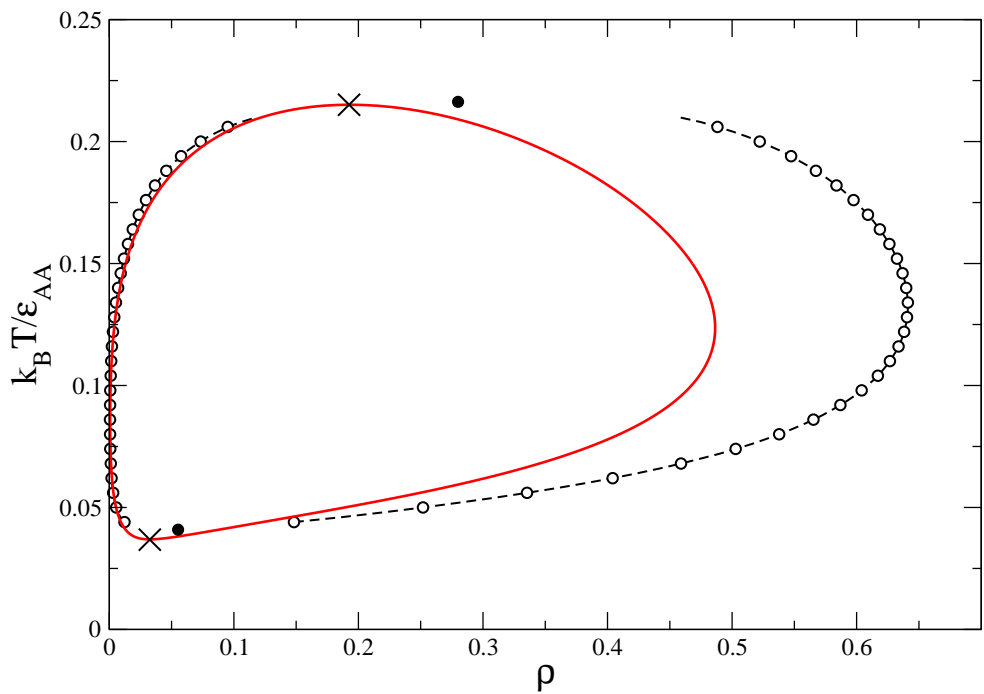}
\caption{Temperature-density phase diagram for a system with $r=0.4$ and $\theta = 120^{\circ}$.
The continuous line represent the theoretical results,  and the points and dashed
lines the results from simulations. Critical points are represented with crosses
(theory) and filled circles (simulation).}
\label{fig:phdeAB04120}
\end{figure}

We start by discussing the results for the model with $\theta=120^\circ$. In figure \ref{fig:phdeAB04120} we plot the phase diagram of the system with $r=0.4$. There are several interesting features, similar to what was found in \cite{rovigatti2013}.
The phase diagram exhibits a closed miscibility loop, with one upper and one lower critical point. Close to the lower critical point, both phases are re-entrant, i.e., the density of the coexisting liquid decreases and the density of the coexisting vapor increases upon cooling. There is semi-quantitative agreement between the upper and lower critical temperatures obtained from the theory and simulation, and qualitative agreement between the density of the liquid binodal, which is underestimated by the theory (in line with the results of previous works).
 
\begin{figure} 
\includegraphics[width=150mm,clip=]{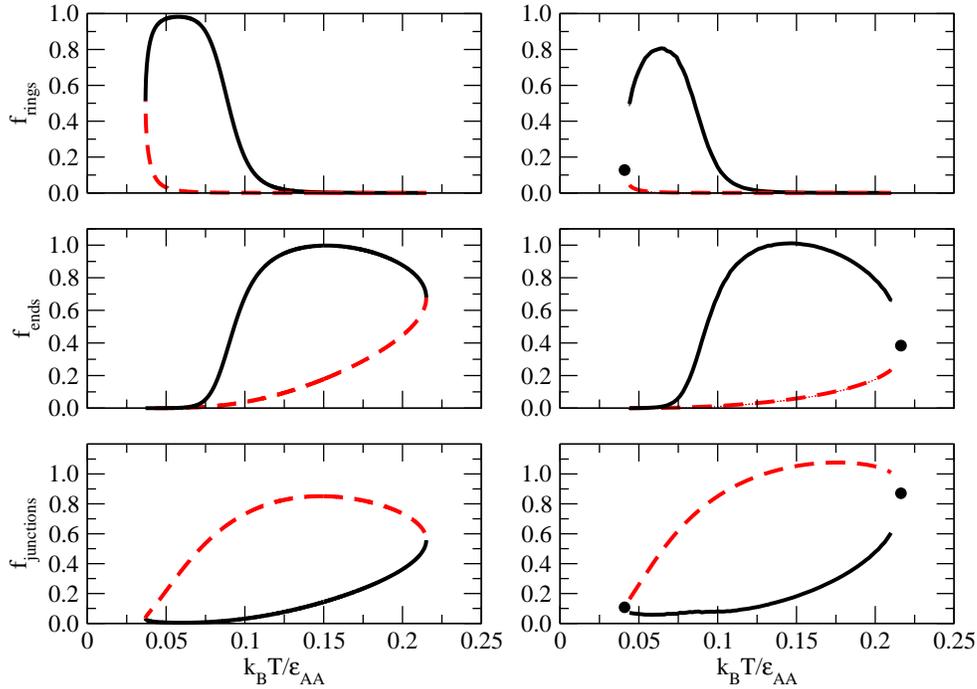}
\caption{Structure of the coexisting phases for a system with $r=0.4$ and $\theta=120^\circ$, as measured by the quantities $f_{rings}$, the fraction of particles in rings, $f_{ends}$, the mean number of unbonded $A$ patches per particle, and $f_{junctions}$, the mean number of bonded $B$ patches per particle. The figures on the left depict the results from the theory and those on the right the results from simulations. The full line corresponds to the vapor phase and the dashed line to the liquid phase.}
\label{fig:struct04120}
\end{figure}

In order to characterize the structure of the coexisting phases, we calculate 3 quantities at coexistence:
(i) the fraction of particles in rings, $f_{rings}\equiv G_1/\rho$;
(ii) the number of ends or unbonded $A$ patches per particle, $f_{ends}\equiv 2X_A$; and (iii) the 
number of junctions or bonded $B$ patches per particle, $f_{junctions}\equiv n(1-X_B)$. 
As shown in figure \ref{fig:struct04120}, close to the upper critical point (at temperatures down to $k_B T/\epsilon_{AA}\approx 0.15$) coexistence is obtained between two phases with practically no rings; the liquid is rich in junctions and poor in ends (high $f_{junctions}$ and low $f_{ends}$) and the vapor is rich in ends and poor in junctions 
(low $f_{junctions}$ and high $f_{ends}$). Therefore, coexistence is between a low density phase of short chains and a high density phase of long chains connected by junctions (a network liquid). Upon cooling, however, the structure of the vapor phase changes dramatically: $f_{rings}$ increases and reaches $\approx 1$, while $f_{ends}$ and $f_{junctions}$ decrease to $\approx 0$; at $k_B T/\epsilon_{AA} \approx 0.05$ the vapor is in practice an ideal gas of rings. On the other hand, the liquid phase retains the network structure, albeit sparser, with longer chains (decreasing $f_{ends}$) and fewer junctions (decreasing $f_{junctions}$). The number of rings in the liquid phase continues to be, at these intermediate temperatures, negligible. 
Finally, close to the lower critical point, $f_{ends}\approx 0$ in both phases. This means that the number of unbonded $A$ patches is negligible, and both phases evolve (on cooling) to a fully connected network of rings and chains (connected by junctions): the liquid has more chains and fewer rings than the vapor. 
  
An interesting feature of the phase behavior (revealed by theory) is the inversion of the usual sign of entropy and energy variation between the coexisting phases. In ordinary liquid-vapor coexistence, the liquid has a lower energy and a lower entropy than the vapor: this happens close to the upper critical point. However, close to the lower critical point, these relations are reversed: the vapor, being, as described earlier, a gas of rings, has a lower energy and a lower entropy than the liquid. 

\begin{figure} 
\includegraphics[width=80mm,clip=]{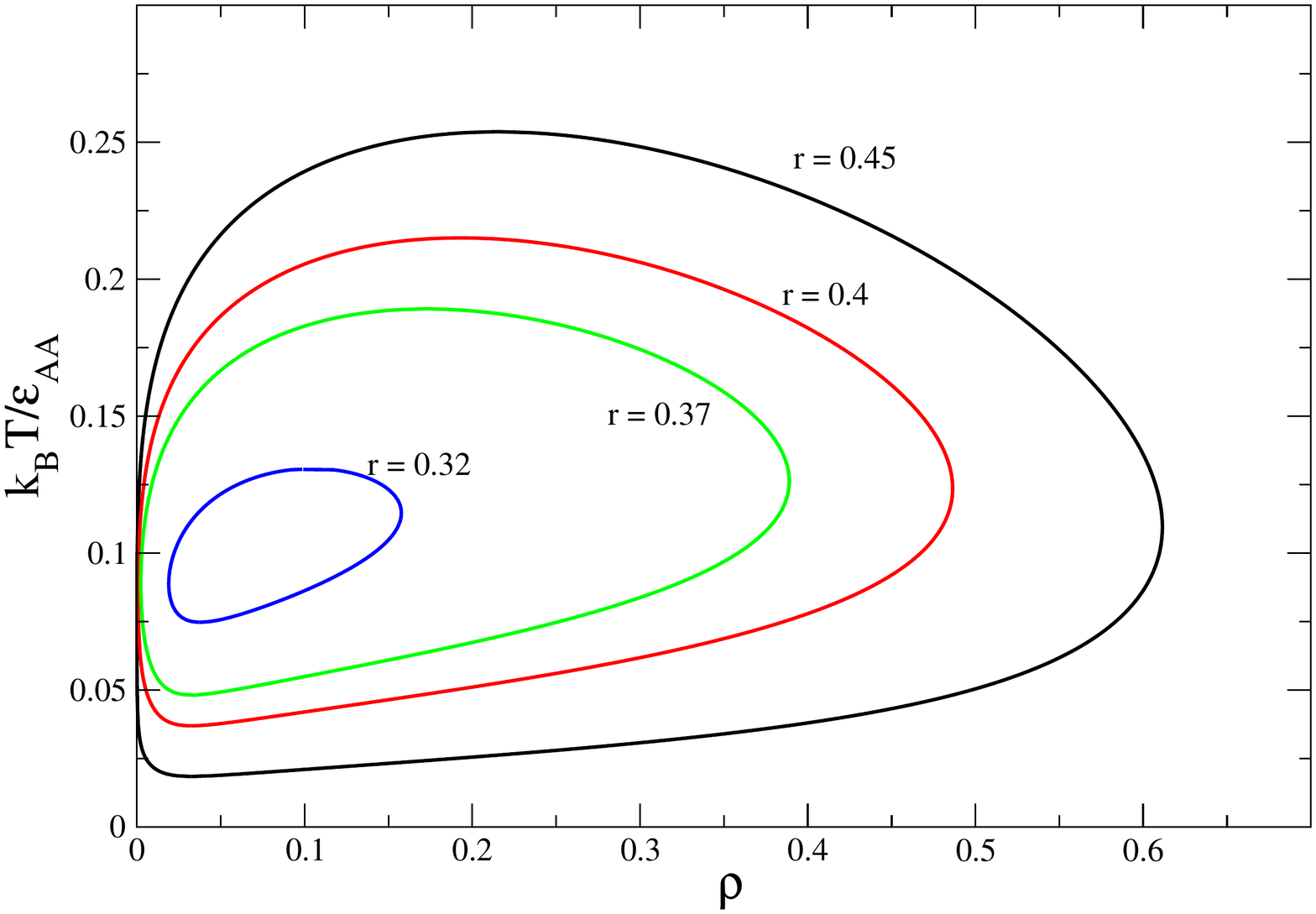}
\includegraphics[width=80mm,clip=]{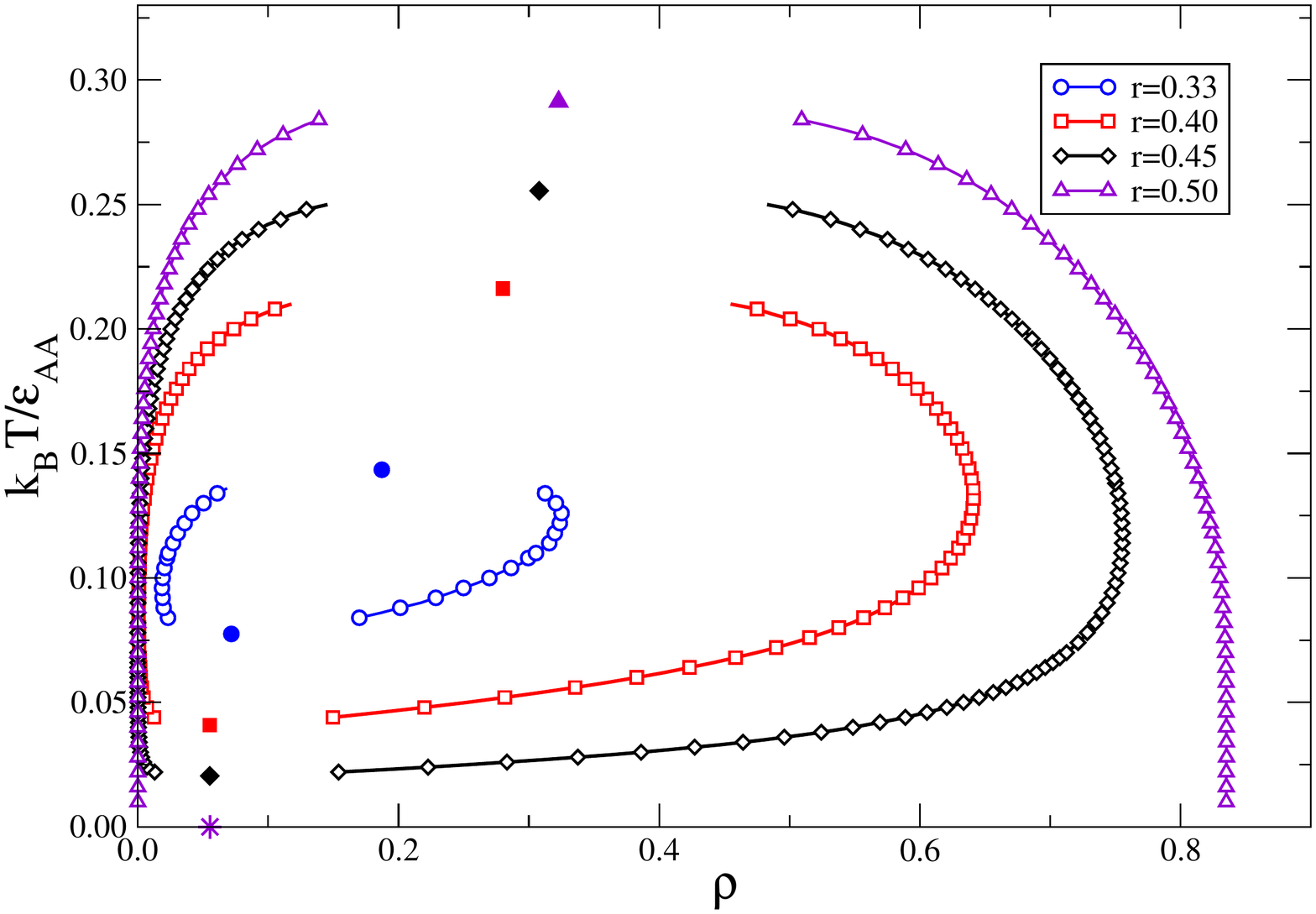}
\caption{Temperature vs density phase diagram for several realizations of the model with $\theta=120^\circ$ and the indicated values of $r$.
Left: theory. Right: simulation; 
Filled symbols mark the location
of the critical points. The asterisk indicates the density of the lower critical point when $ r \rightarrow 0.5^-$}
\label{fig:phd120}
\end{figure}

\begin{figure} 
\includegraphics[width=150mm,clip=]{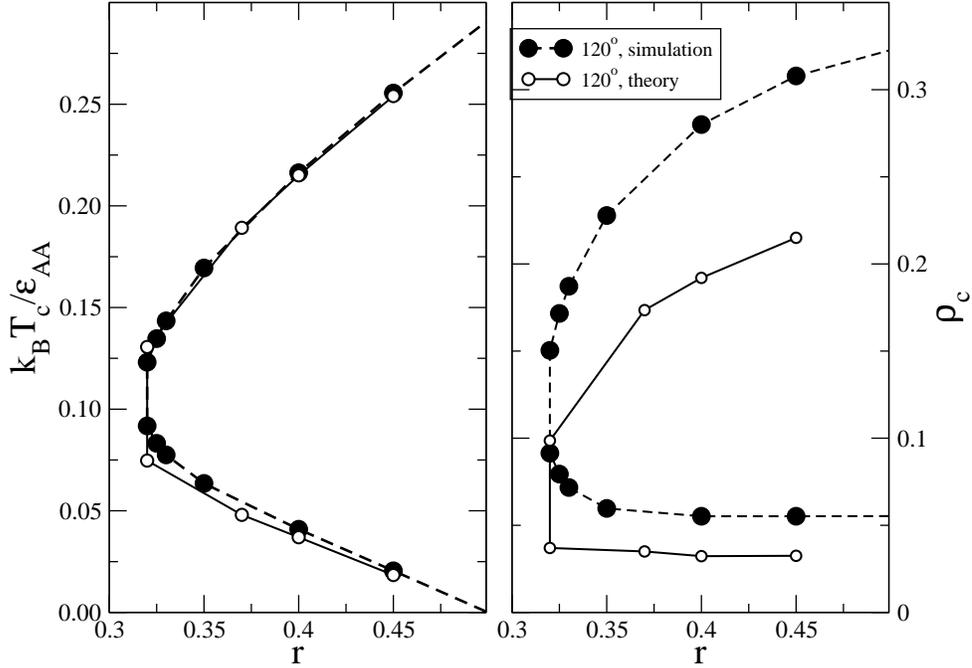}
\caption{Critical temperatures (left) and critical densities (right) for systems with $\theta=120^\circ$ and several values of $r$. Full/open  symbols represent simulation/theoretical results. For a given value of $r$, the larger critical density corresponds to the upper critical point, both in theory and simulation.}
\label{fig:critprop120}
\end{figure}

In order to investigate the effect of decreasing $r$, the binodals for systems with different values of $r<0.5$ were calculated (see figure \ref{fig:phd120}). 
As $r$ decreases, the closed miscibility loop shrinks, and coexistence is obtained in narrower ranges of both temperature and density. 
In fact, there is a threshold value $r^*$ below which no phase separation is found, and self-assembly becomes the only mechanism of aggregation 
(both theory and simulation suggest $r^*\approx 0.3$ for a system with  $\theta=120^\circ$ - see figure \ref{fig:critprop120} and
table \ref{table:1}-). 
This calculation also confirms that it is the energy cost of the junctions or $AB$ bonds that drives phase separation and thus the critical points: 
if this cost is too high, no phase separation occurs. 
The value of $r^*$ (i.e. the thresholds for the energetic cost of junctions) will depend on $\theta$ (see the next subsection) and, 
presumably, on $n$ (the number of $B$ patches). 
We have verified that the features of the phase diagram for $r=0.4$ revealed by figures \ref{fig:struct04120} 
and by the inversion of the usual sign in entropy and energy differences between the phases, are also present in systems with larger values of $r$ in the empty fluid regime, $r^*<r<0.5$.

In figure \ref{fig:critprop120}, the temperatures and densities of both critical points for systems with $\theta=120^\circ$ are plotted as a function of $r$. 
It is clear that in the empty fluid regime, $r<0.5$, 
two types of phase behaviour are obtained:  a closed miscibility loop for $r>r^*\approx 0.3$ and no phase separation for $r<r^*$. 
The agreement between theory and simulation for the critical temperatures is remarkable as noted earlier. 
Both theory and simulation suggest that the lower critical temperature tends to 0 as $r\to 0.5$. Actually, the simulation algorithms developed
in this work allowed us to attain reliable results at very low temperatures, and to simulate the asymptotic cases: $(r \rightarrow 0.50^-, T\rightarrow 0$).
The theory underestimates the critical densities although both theory and simulation predict a density for the lower critical point that is almost constant, for values of $r$ that are not too low, (See Table \ref{table:1})
and smaller than the density of the upper critical point. The latter increases with $r$.    

\subsection{$\theta=90^\circ$,$\theta=60^\circ$}

\begin{figure}
\includegraphics[width=120mm,clip=]{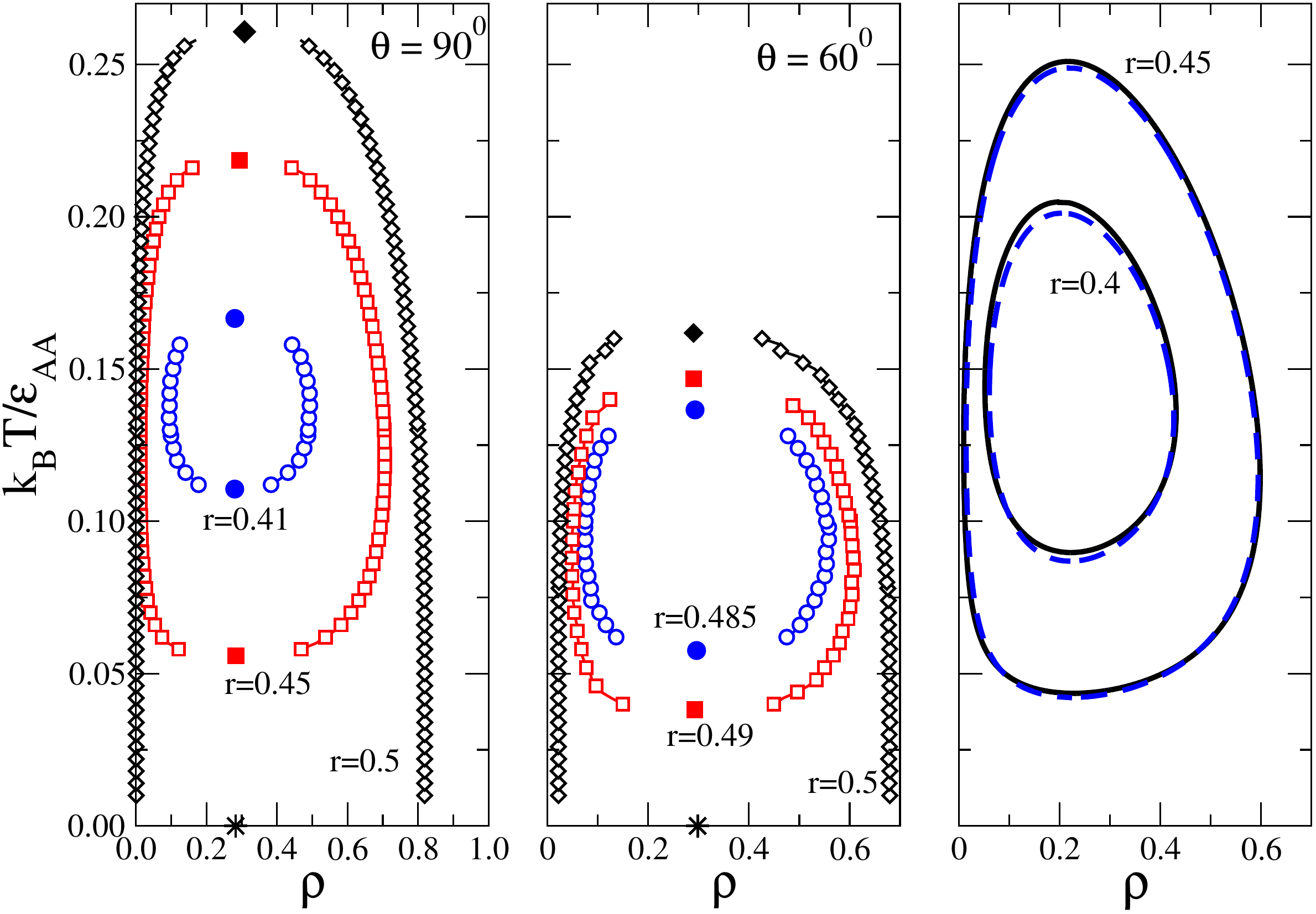}
\caption{Simulation (left and center) and theoretical results (right) for the temperature vs density phase diagram of models with
$\theta=90^{\circ}$ and $\theta=60^{\circ}$ and the indicated values of $r$. In the left and central panels, filled symbols mark the location
of the critical points and the asterisk indicates the density of the lower critical point when $ r \rightarrow 0.5^-$. In the right panel, full lines correspond to $\theta=90^\circ$ and dashed lines to 
$\theta=60^\circ$. Notice that the theory barely distinguishes between the two values of $\theta$.}
\label{fig:phd9060s}
\end{figure}

\begin{figure} 
\includegraphics[width=150mm,clip=]{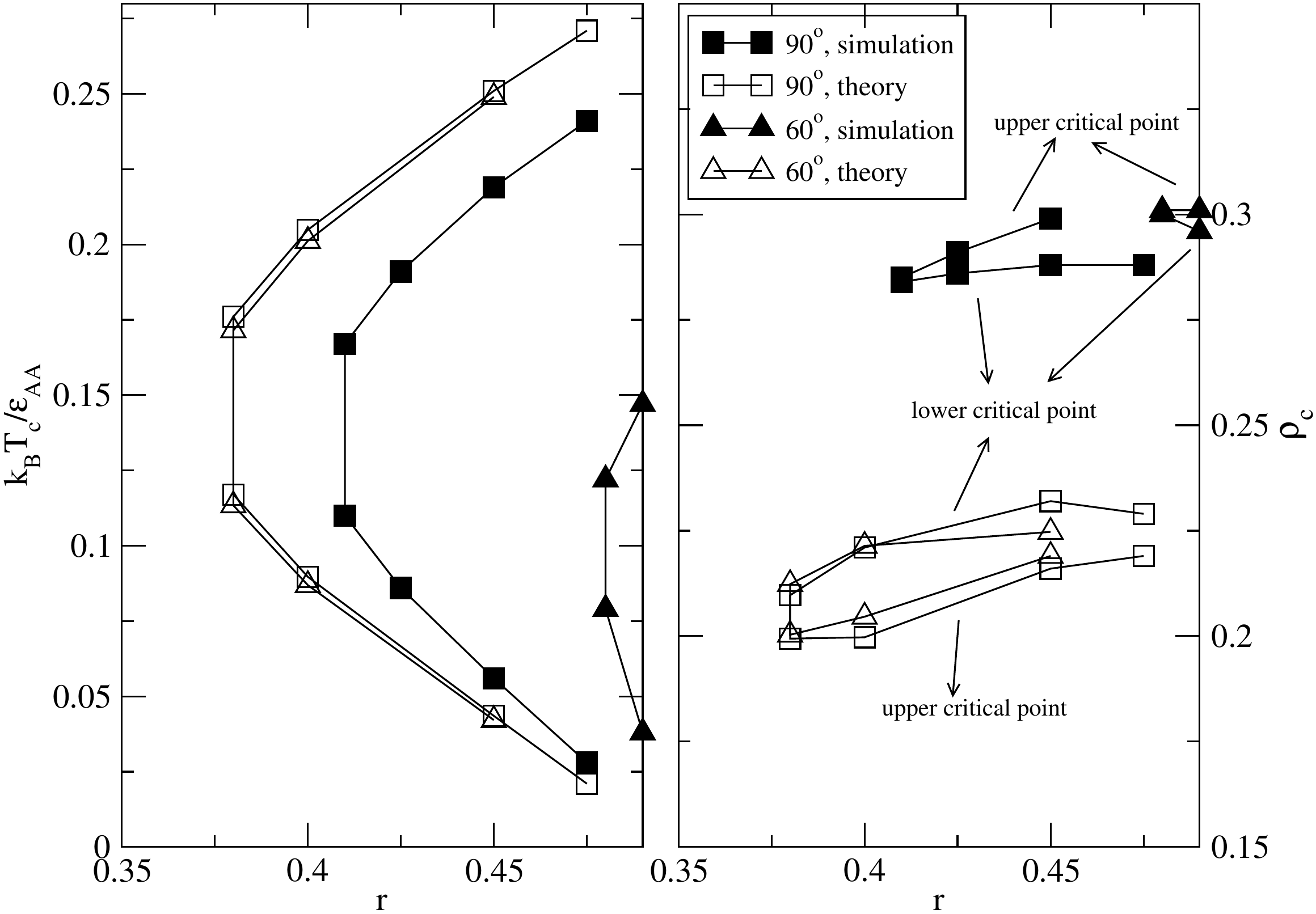}
\caption{Critical temperatures (left) and critical densities (right) as a function of $r$ for systems with $\theta=90^\circ$ (squares) and $\theta=60^\circ$ (triangles). Full/open  symbols represent simulation/theoretical results. For a given value of $r$, the larger density corresponds to the upper critical point in the simulation, and to the lower critical point in the theory.}
\label{fig:critprop6090}
\end{figure}

For systems with $\theta=90^\circ$ and $\theta=60^\circ$, theory and simulation predict the same type of phase behaviour: 
when $r^*<r<0.5$ the binodal consists of a closed miscibility loop and the phase separation is re-entrant close to the 
lower critical point (see figure \ref{fig:phd9060s}).
As shown in figure \ref{fig:critprop6090}, for $\theta=90^\circ$, the theory underestimates the value 
of the threshold $r^*$ and the agreement between the critical temperatures is merely qualitative. 
Both theory and simulation predict upper and lower critical densities, at a given $r$, that are close to each other and that are almost constant as $r$ changes. 
In the simulations the density of the upper critical point is slightly larger than the density of the lower critical point, 
while in the theory the opposite trend was observed. Nevertheless the densities of the upper and lower critical points are not very different
in the range $r^* < r  \le 1/2$, and hardly depend on $r$, as illustrated  
in figure \ref{fig:phd9060s}.

The theory appears to fail badly for systems with $\theta=60^\circ$: the results of figure \ref{fig:critprop6090}
reveal that the theory does not distinguish between this system and that with $\theta=90^\circ$, as almost identical 
results are obtained for the critical temperatures and densities in both cases.
The simulation results reveal that
the range of $r$ where liquid-vapor equilibrium is found is much smaller for $\theta=60^{\circ}$ than for the
other two cases. 

In order to investigate the origin of this discrepancy, we calculate, both by theory 
(using the laws of mass action (\ref{lma1},\ref{lma2},\ref{lma3})) and simulation, several structural quantities as a function of density and temperature, for the family of models under study. 

\begin{figure} 
\includegraphics[width=150mm,clip=]{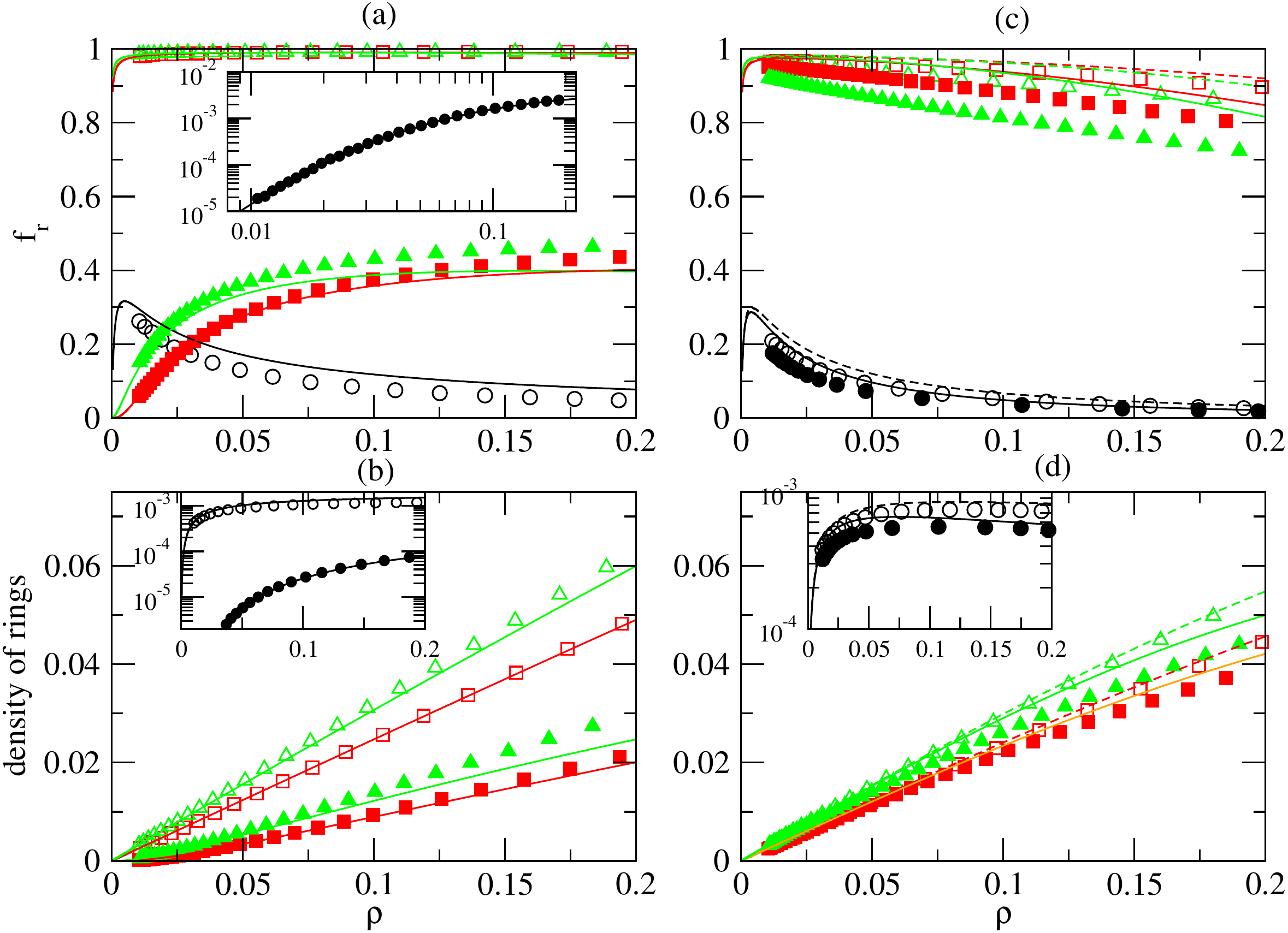}
\caption{Fraction of particles in rings ((a) and (c)) and density of rings ((b) and (d)) 
 as a function of the density. 
Symbols 
represent simulation results for  
$120^o$ 
(circles), $90^o$ (squares) and $60^o$ (triangles), and lines represent the results from theory. 
 In (a) and (b): $r=0$ (no junctions or AB bonds), 
and $k_BT/\epsilon_{AA}=0.1$ (open symbols) and $0.2$ (full symbols).
In (c) and (d): $k_BT/\epsilon_{AA}=0.1$, $r=0.25$ (opens symbols and dashed lines) and $r=0.3$ (full symbols and full lines).}
\label{frG0}
\end{figure}

The comparison between theory and simulation results for the fraction of particles in rings $f_r$ and for the density of rings $G_0$ shown in Figure \ref{frG0}, tests the approximation of the partition function of rings, for all densities and temperatures, by that of a single isolated ring. 
In general, it is seen, as expected, that there is qualitative agreement between theory and simulation, which becomes quantitative at low densities.
Figures \ref{frG0}a) and b) depict results obtained at temperatures $k_{\rm B}T/\epsilon_{AA}=0.1$ and $0.2$ when no $AB$ bonds are present 
($r=0$); figures \ref{frG0}c) and d) display results for $r=0.25$ and $r=0.3$ at $k_{\rm B} T/\epsilon_{AA}=0.1$.
Figure \ref{frG0}a) reveals that there is good agreement between theory and simulation when $f_r$ is either close to 1 or close to 0. For intermediate values of $f_r$ larger deviations are found at lower values of $\theta$: 
the theory underestimates $f_r$ for $60^\circ$ and $90^\circ$, and slightly overestimates it for $120^\circ$.
On the other hand, the results of figure \ref{frG0}b) confirm general good agreement between theory and simulation for $G_0$. Again, larger deviations are found at lower $\theta$: for $\theta=60^\circ$ it is clear that the theory underestimates the density of rings.
The effect of the $AB$ interactions is illustrated in figures \ref{frG0}c) and d). The agreement of both $f_r$ and $G_0$ for $\theta=120^\circ$ is similar to that found for $r=0$. For systems with $\theta=60^\circ$ and $90^\circ$, however, the agreement at high $f_r$ when $r=0$ is lost: the theory overestimates $f_r$ and larger deviations are found at larger values of $r$ and lower values of $\theta$. The same trends are observed in the comparison of $G_0$ shown in figure \ref{frG0}d).

\begin{figure} 
\includegraphics[width=150mm,clip=]{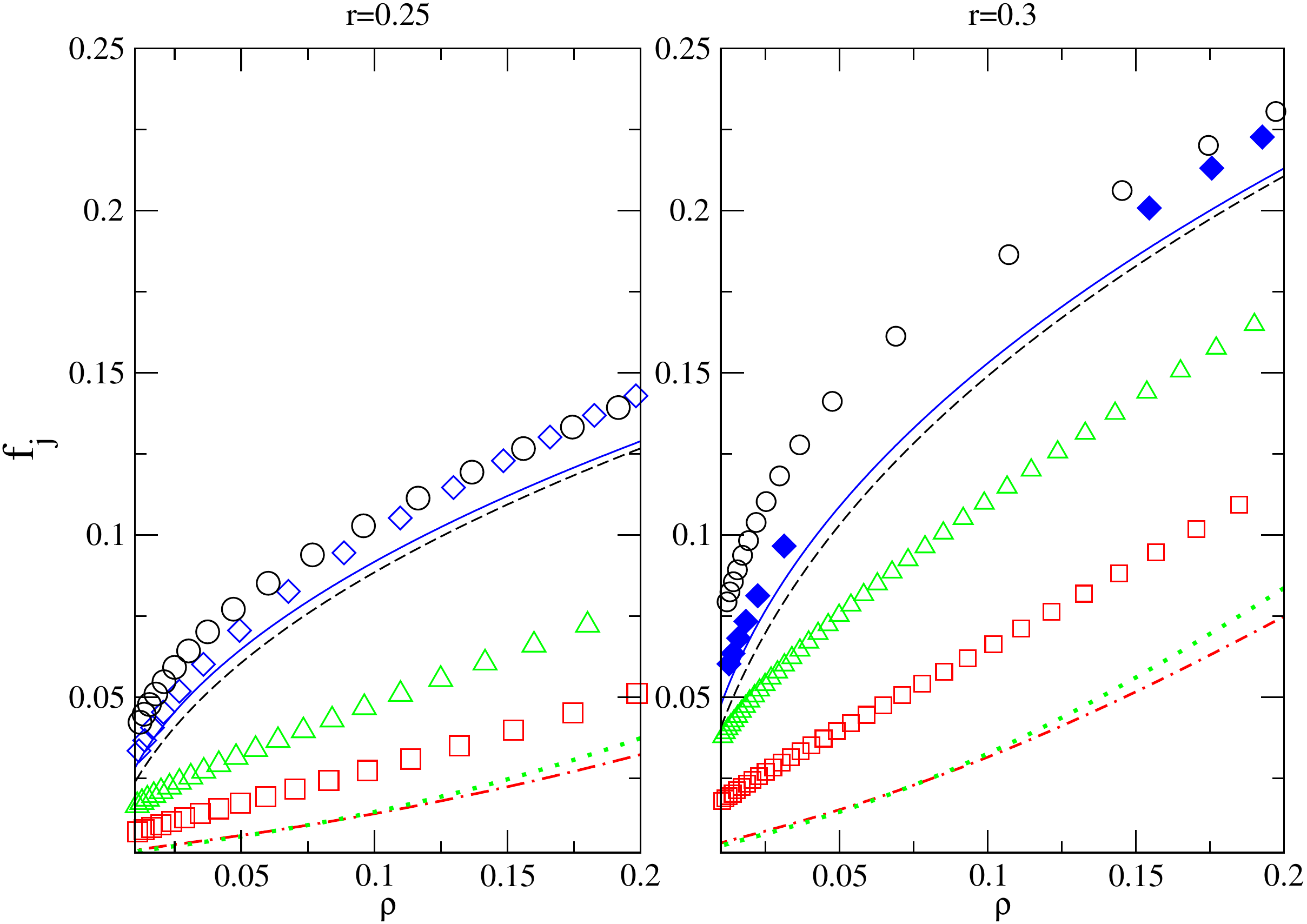}
\caption{Number of junctions (or $AB$ bonds) per particle, $f_j$, as a function of density at $k_BT/\epsilon_{AA}=0.1$. Left panel: $r=0.25$; right panel $r=0.3$. Symbols and lines represent simulation and theoretical results respectively, for different values of $\theta$:
triangles and dotted lines, $\theta = 60^\circ$; squares and dashed - dotted lines, $\theta=90^\circ$; circles and dashed lines, $\theta=120^\circ$; 
diamonds and full lines, $\theta=180^\circ$ (no rings).
}
\label{fjrho}
\end{figure}

Figure \ref{fjrho} displays the theoretical and simulation results for the number of 
junctions (or $AB$ bonds) per particle, $f_j$, as a function of density at $k_BT/\epsilon_{AA}=0.1$,  for
$r=0.25, 0.30$ and for $\theta=60^\circ, 90^\circ, 120^\circ, 180^\circ$. In all cases, we found that the theory underestimates the number of junctions and that the disagreement between theory and simulation increases when $r$ increases and/or 
$\theta$ decreases. This deviation is even observed for $\theta=180^\circ$, where there are no rings \cite{almarza2012}; given that in similar off lattice models \cite{tavares2010} better agreement was obtained, part of the observed discrepancies may be  traced to the correlations introduced by the lattice structure. 
It is also worthwhile to note that for systems with $\theta=120^\circ$ and $180^\circ$ the theoretical and simulation results show a tendency to coincide at low densities. However, for $\theta=90^\circ$ and $60^\circ$, large discrepancies between theory and simulation persist at low densities. 

The deviations between the theoretical and simulation results for $f_r$ and for $f_j$ are consistent:
the theory underestimates the number of $AB$ bonds and thus allows the formation of more rings. This means that the competition between ring formation and branching is not captured quantitatively by the theory. Therefore, the structure predicted by the theory, for systems with $\theta=90^\circ$ and 
$\theta=60^\circ$, is different from that of the simulations and as a result strong discrepancies in the phase behaviour are found. 

\section{Conclusions}
\label{sec:Conclusions}

We have shown, through the investigation of a patchy particle model on fcc lattices, both theoretically and by simulation, that ring formation promotes the appearance of close miscibility loops, with two critical points, as the result of self-assembly of a single component system where linear structures are energetically favored and branching is entropically favored (but energetically unfavored).  
Near the upper critical point, the coexisting phases consist of a vapor of short chains and a network liquid (long branched chains). On the other hand, close to the lower critical point, the low density phase is a vapor of rings and the high density phase a network of chains and rings.
We have performed a systematic study of the phase behavior of this model, by varying the relevant parameters: $r\equiv\epsilon_{AB}/\epsilon_{AA}$, the energy of branching (relative to that of chaining) and 
$\theta$, the angle between the directions of the $A$ patches in a particle, which determines the entropic cost of forming rings.   
Closed loops were found, generically, in a range of bonding energies, $r^*(\theta)<r<1/2$. The
threshold $r^*(\theta)$ increases as $\theta$ decreases: if shorter rings are promoted (by decreasing $\theta$), the energy cost of branching has to decrease (i.e. $r$ increases) to obtain phase separation. It was also found that, for $r<r^*(\theta)$, there is no phase separation.

The general results are obtained both by theory and simulation. However, quantitative agreement between the two deteriorates rapidly as $\theta$ decreases, from almost quantitative for $\theta=120^\circ$ to merely qualitative for $\theta=60^\circ$.    
We have also found that, as $\theta$ decreases, the theory tends to overestimate the fraction of particles in rings and to underestimate the fraction of junctions (or AB bonds) per particle. These deviations become more pronounced as $r$ increases. 

The origin of the discrepancy may be traced to the way in which the AB bonds are accounted for in the theory. 
It is known \cite{jackson1988,Bianchi2008} that the thermodynamics and structure that results from WPT
is equivalent to that of an ideal mixture of tree like clusters: loops are absent and each bond formed decreases the number of clusters by one. 
The only loops considered in the generalization of WPT used in this work are rings formed by closed sequences of $AA$ bonds: the formation of an AB bond is still assumed (as in the original WPT) to decrease the number of clusters by one. 
The different treatment given to the $AA$ rings is justified by their lower energy when compared with loops of the same size 
that contain $AB$ bonds. However, when $\theta$ is decreased, the entropic cost of forming a loop with $AB$ bonds decreases; one then expects formation of more loops of this type and an eventual failure of the approximation.  

The results of simulations seem to corroborate this idea. In fact, in systems where $AA$ rings are absent and loops are difficult to form (i.e. when $\theta=180^\circ$) \cite{almarza2012}, the possibility of forming $AB$ bonds correspond to a decrease in $r^*$. This happens since almost all of these bonds lead to branching of chains and not to loop formation. 
By contrast, in the work reported here, we found that, by decreasing $\theta$, $r^*$ increases and that the number of $AB$ bonds also increases.  
Therefore, at low $\theta$, most $AB$ bonds form loops; phase separation is driven by the $AB$ bonds that lead to branching, and to form these in sufficient numbers, their energy cost has to decrease (i.e. $r^*$ has to increase, as observed in the simulations). 
A quantitative investigation of the effect of loops with $AB$ bonds in the phase behaviour of this family of models will be addressed in future work.

\section*{Acknowledgments}
J.M.T. and M.M.T.G. acknowledge financial support from the Portuguese Foundation for Science and Tecnhology under contract EXCL/FIS-NAN/0083/2012 and PEst-OE/FIS/UI0618/2011.
NGA acknowledges the support from the Direcci\'on
General de Investigaci\'on Cient\'{\i}fica  y T\'ecnica under Grant
No. FIS2010-15502 and from the Direcci\'on General de
Universidades e Investigaci\'on de la Comunidad de Madrid under
Grant No. S2009/ESP/1691 and Program MODELICO-CM.

\end{document}